\documentclass[10pt,conference]{IEEEtran}
\usepackage{amsmath}
\usepackage{amsfonts}
\usepackage{amssymb}
\usepackage{graphicx}
\usepackage{epstopdf}

\usepackage{array}

\newcommand*{\vertbar}{\rule[-1ex]{0.5pt}{2.5ex}}

\begin{document}
%
%
\title{Neural Signal Multiplexing via Compressed Sensing}
\author{
\IEEEauthorblockN{Nithin Nagaraj\IEEEauthorrefmark{1} and K. R. Sahasranand\IEEEauthorrefmark{2}}
     \IEEEauthorblockA{\IEEEauthorrefmark{1}Consciousness Studies Programme, School of Humanities, National Institute of Advanced Studies, \\ Indian Institute of Science Campus, Bangalore, INDIA. \\
     \IEEEauthorrefmark{2}Department of Electrical Communication Engineering,
Indian Institute of Science, Bangalore, INDIA. \\(Email: nithin@nias.iisc.ernet.in, sanandkr@ece.iisc.ernet.in)}
          }
%
%
%
\maketitle
%
\begin{abstract}
Transport of neural signals in the brain is challenging, owing to neural interference and neural noise. There is experimental evidence of multiplexing of sensory information across population of neurons, particularly in the vertebrate visual and olfactory systems. Recently, it has been discovered that in lateral intraparietal cortex of the brain, decision signals are multiplexed with decision-irrelevant visual signals. Furthermore, it is well known that several cortical neurons exhibit chaotic spiking patterns. Multiplexing of chaotic neural signals and their successful demultiplexing in the neurons amidst interference and noise, is difficult to explain. In this work, a novel compressed sensing model for efficient multiplexing of chaotic neural signals constructed using the Hindmarsh-Rose spiking model is proposed. The signals are multiplexed from a pre-synaptic neuron to its neighbouring post-synaptic neuron, in the presence of $10^4$ interfering noisy neural signals and demultiplexed using compressed sensing techniques.
\end{abstract}
\begin{IEEEkeywords}
neural signal multiplexing, compressed sensing, chaotic signals.
\end{IEEEkeywords}
%
%
%
\section{Introduction}
\par The number of neurons in the human brain is approximately $86$ million~\cite{Azevedo2009} which is roughly half the number of stars in our Milky Way galaxy. Each of these neurons is connected anywhere between $10^2$ to $10^4$ other neighbouring neurons on an average. It remains a mystery to a large extent as to how information encoded in action potentials (spikes) is transported from one neuron in the middle of a labyrinth of neurons to its immediate neighbour amidst a cacophony of interfering neural signals and neural noise. The brain not only employs very efficient compression of information with low consumption of energy, but also is known to robustly multiplex neural signals.

\par In a recent study involving adult rhesus monkeys, single neuron spiking measurements in the lateral intraparietal area (LIP) showed that
 decision signals are multiplexed with decision-irrelevant visual signals~\cite{Meister2013}. Furthermore, the same study also revealed that neurons in LIP exhibit diverse dynamics indicating a broad range of neural computations. Friedrich et al.~\cite{Friedrich2004} found that oscillatory field potentials in the zebrafish olfactory bulb multiplex information about complimentary stimulus features by the same population of neurons. These can be retrieved selectively by biologically plausible mechanisms. Panzeri et al.~\cite{Panzeri2010} present a strong case for temporal multiplexing where neural responses at different timescales encode different stimulus attributes. They argue that such multiplexing increases encoding capacity of neural responses and offers advantages over single response timescales, thus enabling the brain to arrive at an information-rich and stable representation of an otherwise noisy and variable sensory environment.

\par In this paper, we propose a phenomenological model to study the feasibility of such efficient multiplexing of neural signals in the presence of neural interference from a very large number of neurons. We wish to point out the possibility of such mechanisms from a theoretical perspective while trying to use reasonable models which simulate the behaviour of actual neurons. To the best of our knowledge, there are no mathematical models in the neuroscience literature which can explain a mechanism of simultaneous compression, reliable multiplexing and robust demultiplexing of chaotic neural signals (amidst $10^4$ noisy signals) which the brain seems to be performing successfully. We hope to demonstrate via theoretical arguments and simulation results that our proposed compressed sensing model has all the three benefits.

\par The paper is organized as follows. We briefly review existing methods of multiplexing of chaotic signals and point out their drawbacks in Section~\ref{sec:multiplex}. None of these existing methods seem feasible to model multiplexing of chaotic neural signals in the brain. We propose a novel model of multiplexing (and demultiplexing) of chaotic neural signals using the framework of compressed sensing in Section~\ref{sec:proposed}. We conclude with future research directions in Section~\ref{conclusion}.

\section{Multiplexing of Chaotic Signals}
\label{sec:multiplex}
\par Chaos as a possible neuronal code was proposed by researchers in neuroscience owing to enormous experimental evidence of chaos in the nerve cells, neural assemblies, behavioural patterns as well as higher brain functions. There are several advantages of chaos and nonlinear response models over stochastic ones, especially the possibility of control of the former. Korn et al., in their excellent review article~\cite{Korn2003}, trace the history of the search for chaos in the nervous system providing pointers to numerous experimental studies. Dynamical approach to brain operations and cognition has been vindicated and there is a large body of compelling evidence of chaos at all levels.

\par The problem that we wish to address is the ability of neurons to multiplex chaotic signals. To this end, we briefly review known methods for multiplexing of chaotic signals. Starting from $1996$, continuous time chaotic signals have been multiplexed using the notion of dual synchronization~\cite{Multiplex1}. The idea of dual synchronization proposed by Liu and Davis~\cite{Multiplex2} is to use a scalar signal to simultaneously synchronize two different pairs of chaotic oscillators. Dual synchronization has been implemented in electronic circuits~\cite{Multiplex3}, but this has problems such as difficulty of extending to more than two signals, sensitivity to noise and the need to carry two scalar signals for multiplexing more than two signals (but action potentials are transmitted as just a single scalar signal). Even $1\%$ of noise results in a synchronization error of $4\%$ as reported by Liu and Davis~\cite{Multiplex2}.  Symbolic dynamics-based approach to multiplex two pairs of low frequency chaotic electronic circuits that produce R\"{o}ssler-like oscillations was demonstrated by Blakely and Corron~\cite{Multiplex4}. However, even a small synchronization error leads to huge errors thereby making it unsuitable. Nagaraj and Vaidya~\cite{Multiplex5} proposed multiplexing of discrete chaotic signals in the presence of noise using the idea of symbolic sequence invariance. While they demonstrate multiplexing up to $20$ chaotic signals, their method is not robust to huge amounts of noise and as the number of signals to be multiplexed increases, the noise tolerance comes down quite rapidly.

\par Even if we grant that some of the above methods could be modified to work with multiple chaotic signals, it is not at all obvious whether these can be extended to multiplex neural signals from neuronal models, as it has never been attempted before. Most importantly, these methods would fail to address the problem that we are interested in $-$ multiplexing a small number of neural signals amidst $10^4$ competing noisy neural signals from neighbouring neurons. These methods have never been tested for such large number of signals, but this is precisely what the brain has to contend with.
\section{Proposed Approach}
\label{sec:proposed}
\par The problem addressed in this work is stated as follows. Given a particular pre-synaptic neuron ($Z$) and its eastern post-synaptic neighbour ($E$), both of these being connected to $\approx 10^4$ other neurons, how does neuron $Z$ simultaneously transport 4 neural signals $s_1(n)$, $s_2(n)$, $s_3(n)$ and $s_4(n)$ (corresponding to North, North-West, South-West and South neighbours respectively) to neuron $E$? Here $n$ stands for discrete time index (we assume a sampled neural signal, but in what follows, our methods can be extended to continuous time signals as well). The neuron $Z$ can only transmit a single scalar signal across its synapse to neuron $E$. It is to be remembered that all neural signals are always corrupted with some amount of noise, which we shall assume to be additive Gaussian\footnote{More realistic noise models need to be tested in the future.}.
\subsection{Neural Signal Multiplexing}
\label{subsec:mux}
\par At neuron $Z$, up to a first order approximation, we shall assume that it forms a weighted addition of all its input noisy neural signals. This can be mathematically represented as:
\begin{equation}\label{eq1}
  y(n) = \sum_{i=1}^{N} w_i \cdot s_i(n),
\end{equation}
where $n=1, \ldots, M$. Equivalently, the above equation could be expressed in matrix form:
\begin{equation} \label{eq2}
\left[
  \begin{array}{c}
    y(1) \\
        \vdots\\
    y(M)
  \end{array}
\right]
   =
\left[
  \begin{array}{cccc}
    \vertbar & \vertbar &        & \vertbar \\
    s_{1}    & s_{2}    & \ldots & s_{N}    \\
    \vertbar & \vertbar &        & \vertbar
  \end{array}
\right]
\left[
  \begin{array}{c}
    w_1 \\
    w_2   \\
    \vdots\\
    \vdots\\
    \vdots\\
    w_N
  \end{array}
\right],
\end{equation}
which can be further written succinctly as:
\begin{equation}\label{eqYequalAX}
y = Ax.
\end{equation}
Some points to note:
\begin{enumerate}
\item It is quite reasonable to assume linear processing of inputs as in the above equation. For example, the olfactory bulb linearly processes fluctuating odor inputs. Gupta et al.~\cite{Gupta2005} have shown that individual mitral/tufted cells (in anesthetized rats) sum inputs linearly across odors and time.
\item In this work, we assume a value of $N=10^4$. It may be the case that there are fewer than $10^4$ neurons connecting to $Z$ (and $E$). We have assumed this huge number as a worst-case scenario, but our methods work with lower values of $N$ as well.
\item  We shall assume that only a small fraction (say $1.5 \%$) of the $N$ neural signals $\{ s_i\}$ reaching $Z$ are from neighbouring neurons. These are chaotic outputs of the Hindmarsh-Rose neuron model (described below) but corrupted with small amounts of additive noise. Hence, this small set of signals ($1.5 \%$ of the columns of matrix $A$) is modeled as signal-dominant with low noise (additive Gaussian noise\footnote{Let $\mathcal{N}(\mu,\sigma^2)$ denote a Gaussian distribution with mean $\mu$ \& variance $\sigma^2$.} $\mathcal{N}(0,0.01)$). The remaining $98.5 \%$ of the neural signals are completely dominated by additive Gaussian noise and are modeled as i.i.d Gaussian samples $\mathcal{N}(0,1)$. Even though these vast number of columns of $A$ are also neural signals, they are effectively modeled as Gaussian noise, since by the time these neural signals arrive at neuron $Z$ (as well as $E$ - remember that $E$ is nearest neighbour of $Z$), they are dominated completely by noise with no signal component whatsoever.
\item The weights $\{w_i\}_{i=1}^{N}$ are scalars (real-valued) and correspond to each pre-synaptic neuron of $Z$ (these weights could be thought of as contributing to either an inhibitory or an excitatory response). These weights are completely unknown to $E$. It is reasonable to assume that only a few neural signals (say $k$) among the $1.5 \%$ signal-dominant columns of $A$ (in our study we have taken $k=4$) need to be multiplexed from $Z$ to $E$. Hence, we shall assume that only this $k$ sparse set of weights are significant (greater than some threshold $T$), though the actual number of significant weights ($k$) itself is unknown at $E$\footnote{It is this sparsity or rather compressibility of $x$ that comes to our rescue.}.
   \end{enumerate}
\subsection*{Hindmarsh-Rose Neuron Model}
\par The Hindmarsh-Rose neuron model~\cite{Hindmarsh1984} is a widely used model for bursting-spiking dynamics of the membrane voltage $S(t)$ of a single neuron. The equations of the model (in dimensionless form) are:
\begin{eqnarray}
  \dot{S} &=& P + 3S^2 - S^3 - Q + I, \\
  \dot{P} &=& 1 - 5S^2 - P, \\
  \dot{Q} &=& -r \large{[} Q - 4(S + \frac{8}{5}) \large{]}.
\end{eqnarray}
\par
\begin{figure}[!hbp]
\begin{center}
\centering
\resizebox{1.0\columnwidth}{!}{
\includegraphics{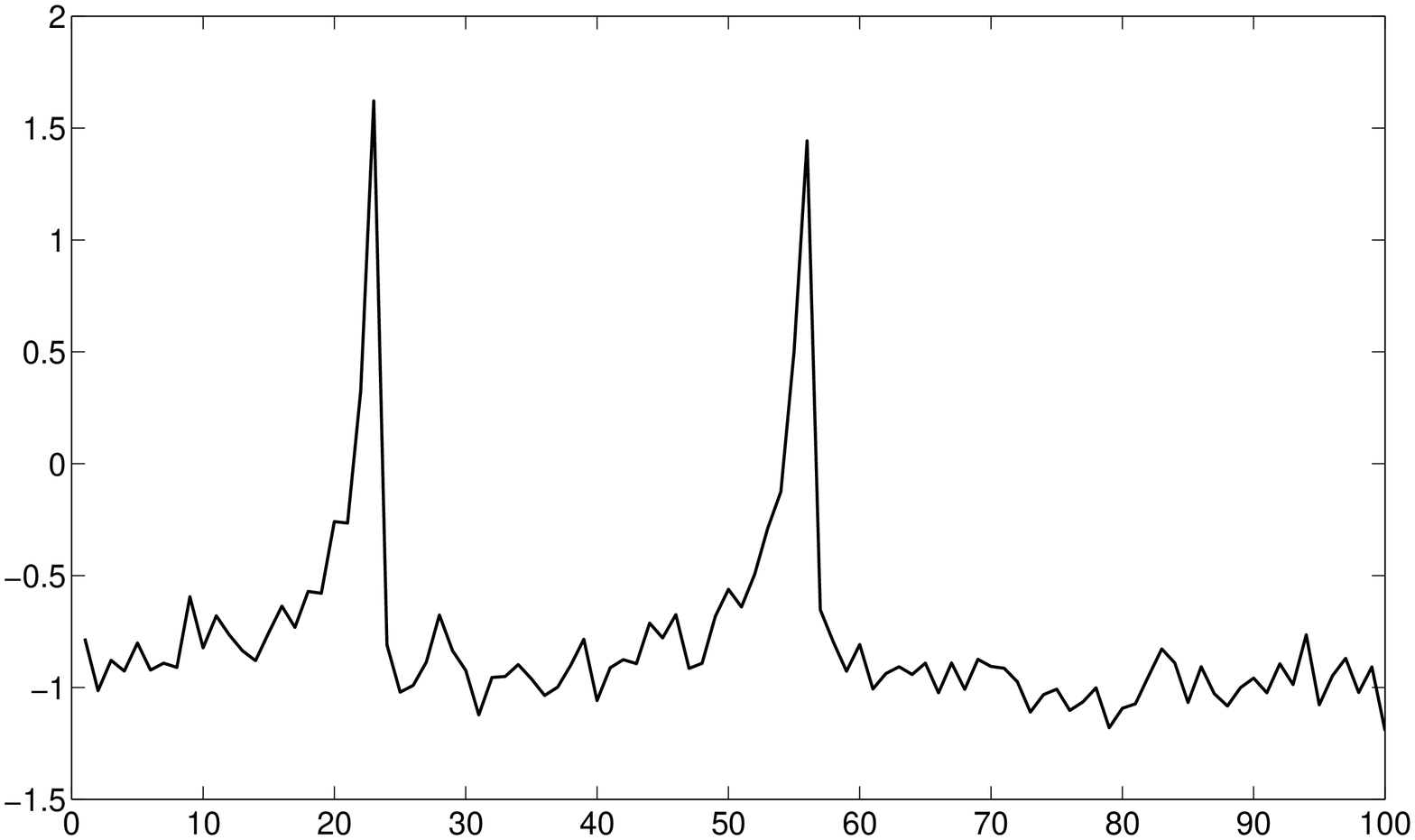}
}
\resizebox{1.0\columnwidth}{!}{
\includegraphics{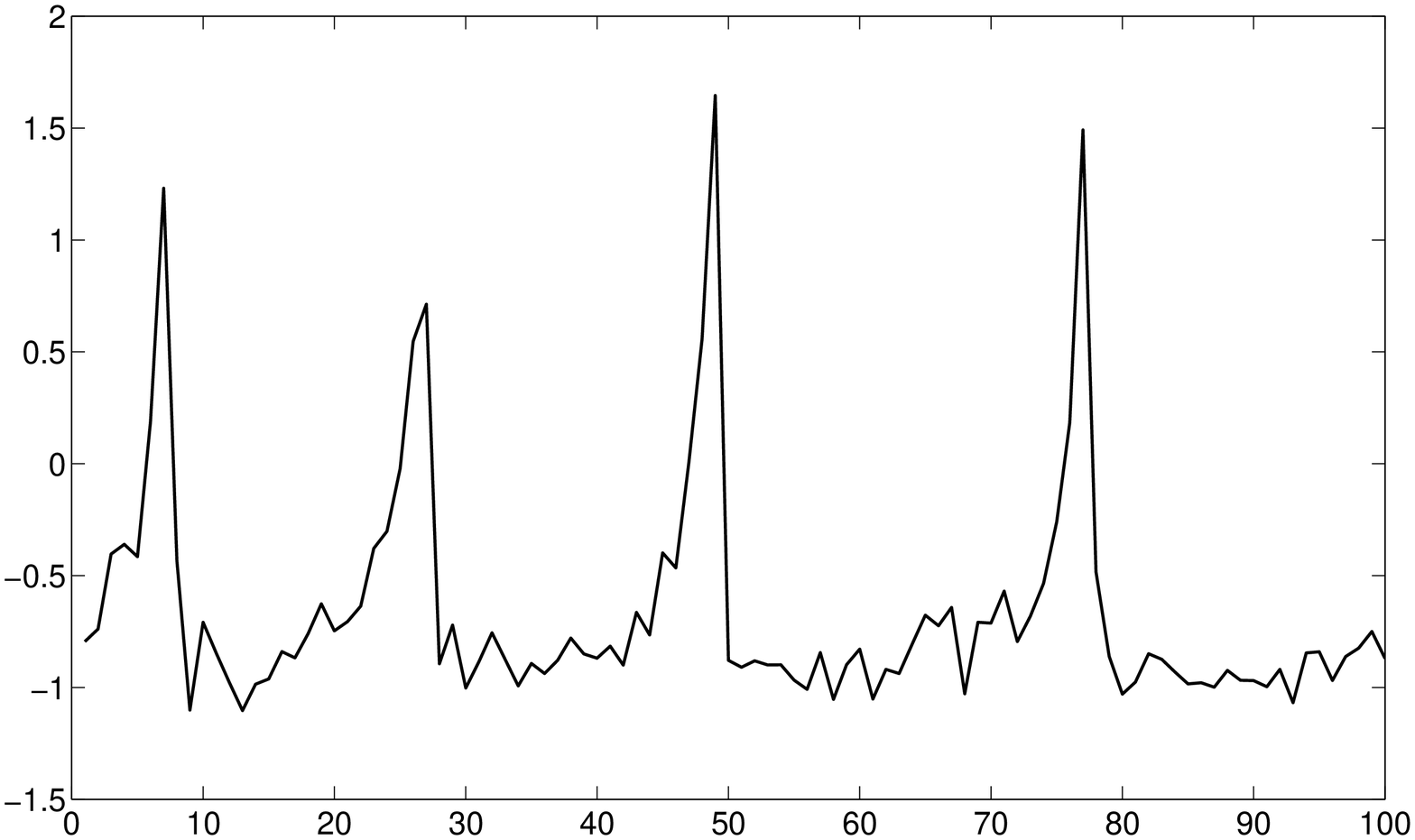}
}
\resizebox{1.0\columnwidth}{!}{
\includegraphics{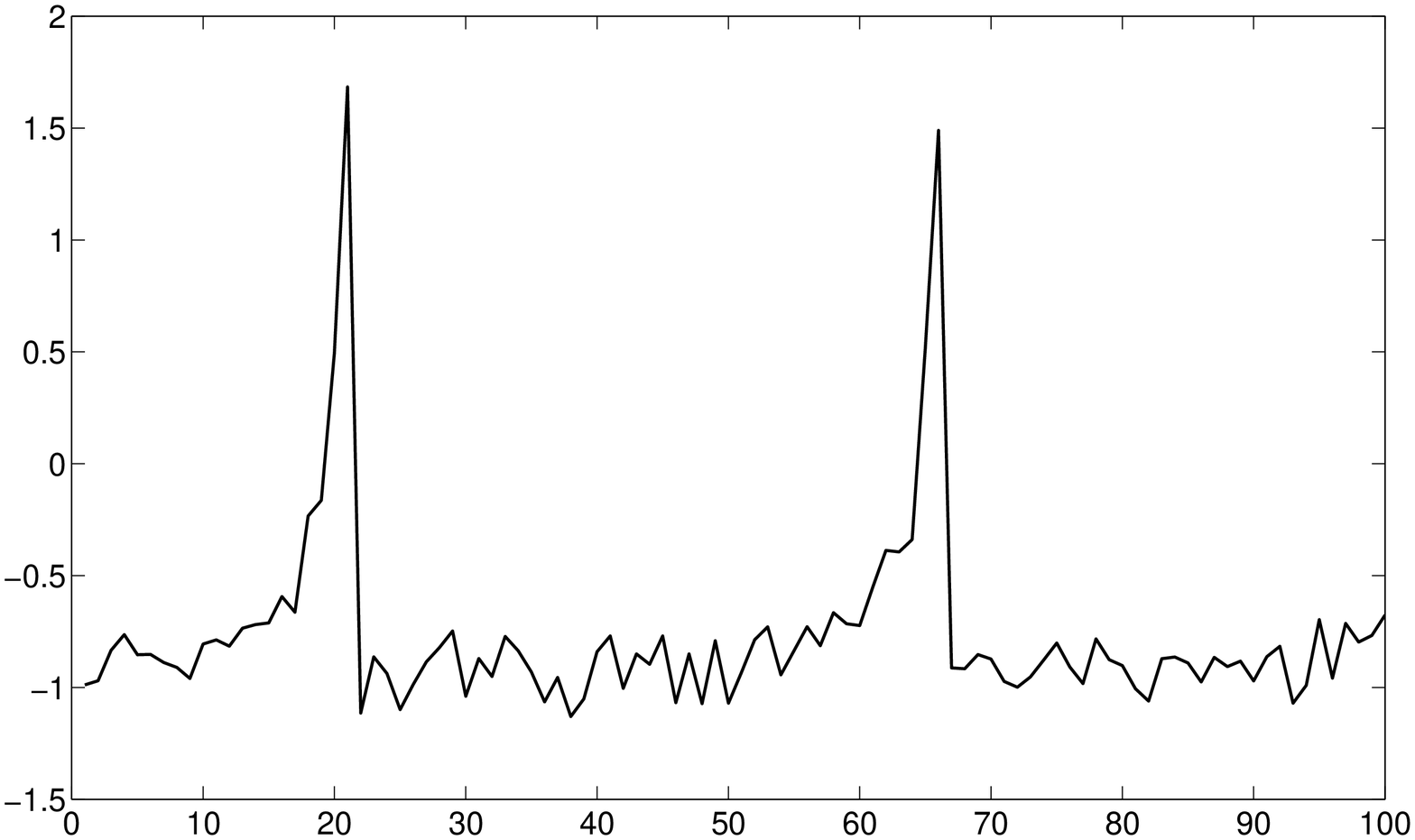}
}
\caption{The three noisy neural signals to be multiplexed along with $s_1(n)$. Top: $s_2(n)$, Middle: $s_3(n)$, Bottom: $s_4(n)$.} \label{fig:signals2}
\end{center}
\end{figure}

\begin{figure}[!h]
\begin{center}
\centering
\resizebox{1.0\columnwidth}{!}{
\includegraphics{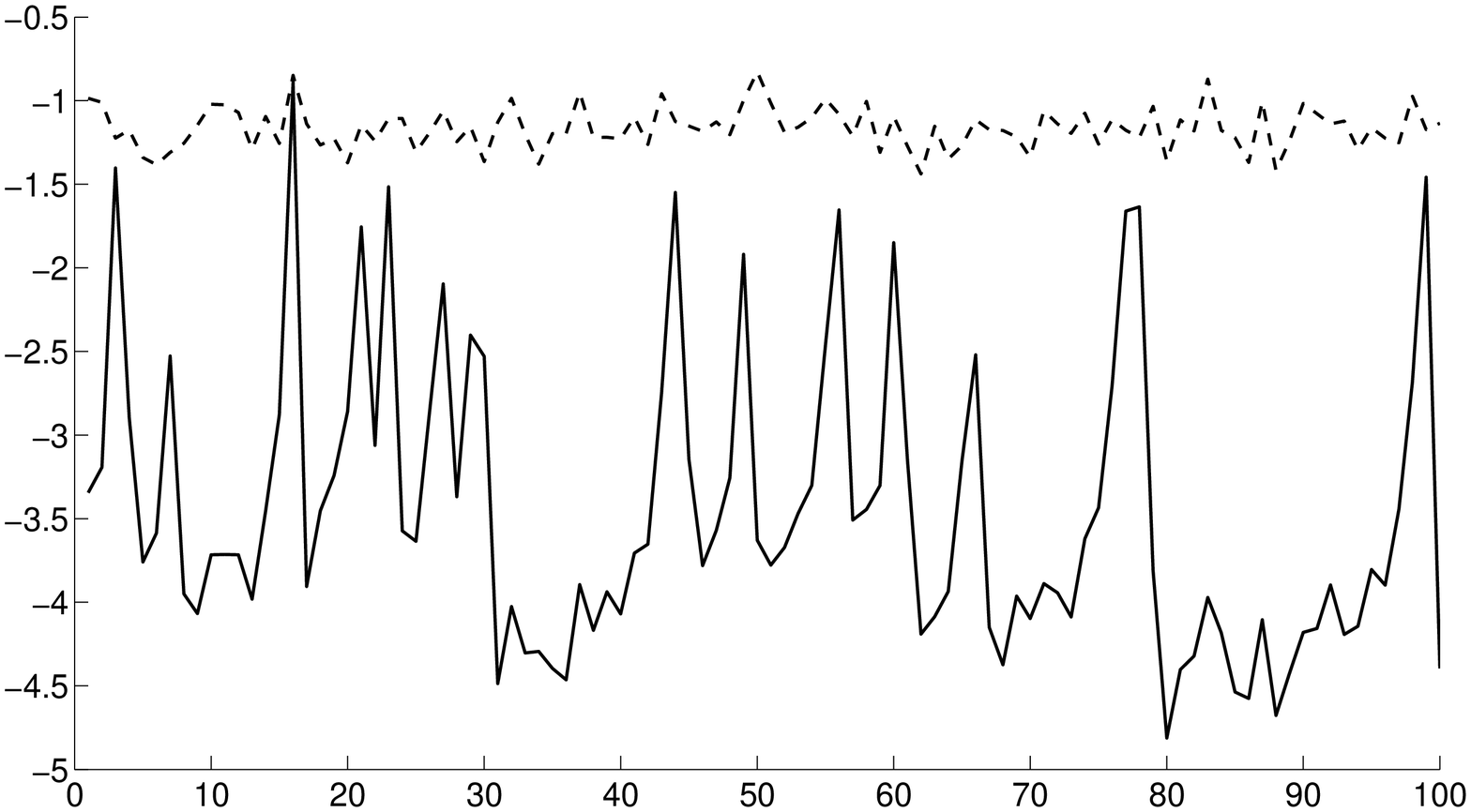}
}
\caption{Signal vs. Noise. The mixed scalar signal $y(n)$ (computed using Equation~(\ref{eq1})) transported from the pre-synaptic neuron $Z$ to post-synaptic neuron $E$ is shown in solid. Neural interference and noise from $9996$ neurons (excluding the effect of the $4$ signals to be multiplexed) is shown in dotted.} \label{fig:ysignal}
\end{center}
\end{figure}

 In the above set of differential equations, $P(t)$ and $Q(t)$ represent auxiliary variables corresponding to fast and slow transport processes across the membrane respectively. The control parameters of the model are the external current applied, $I$, and the internal state of the neuron, $r$. We use a value of $r=0.0021$, as it yields dynamics which corresponds to a realistic description of the electrophysics of axons of certain neurons~\cite{Rabinovich1997}. The model is capable of displaying both regular spiking as well as bursting (chaotic) dynamics for suitable regimes of the parameter $I$. The bursting behaviour exhibits chaos where there is alternating irregular spiking followed by a non-oscillatory phase which suddenly ends in a burst of spikes.
\begin{figure}[!hbp]
\begin{center}
\centering
\resizebox{1.0\columnwidth}{!}{
\includegraphics{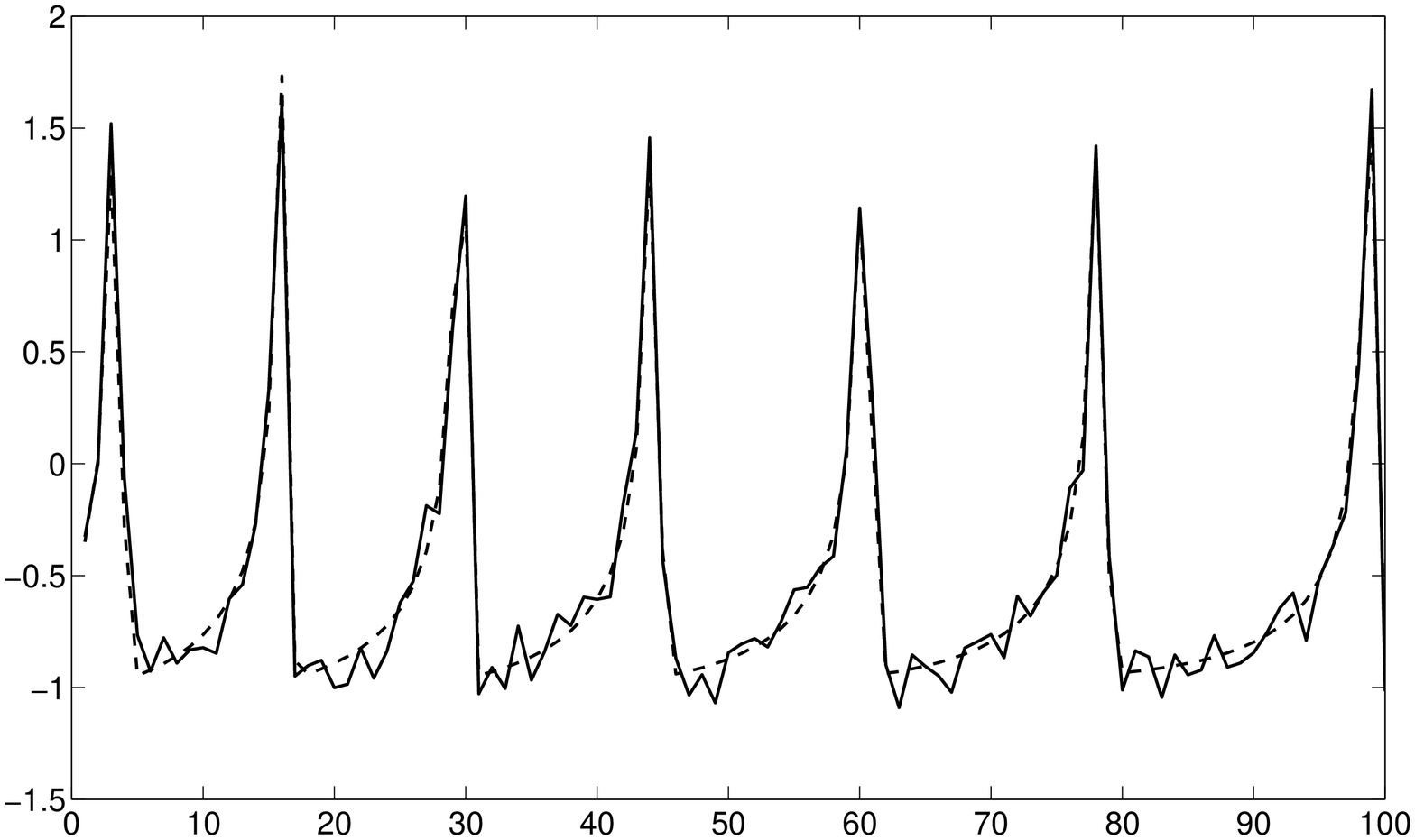}
}
\caption{Noiseless neural signal $s(n)$ simulated using the Hindmarsh-Rose neuron model (dotted). The noisy signal $s_1(n)$ (solid) is obtained by adding $\mathcal{N}(0,0.1)$ to $s(n)$. All such plots in this paper depict signal amplitude against discrete time index $n$.} \label{fig:signal}
\end{center}
\end{figure}
In our study, we have chosen $I=3.28$ which is in the chaotic regime~\cite{Gonzalez2003}. We have sampled $S(t)$ in steps of $1$ time unit to yield discrete-time neural signal $s(n)$ and subsequently add i.i.d. Gaussian noise $\mathcal{N}(0,0.01)$ to yield the final signal $s_1(n)$ as shown in Fig.~\ref{fig:signal}.  In Fig.~\ref{fig:signals2}, we plot the remaining three (noisy) signals $s_2(n)$, $s_3(n)$  and $s_4(n)$ obtained in a similar fashion from randomly chosen initial conditions of the neuron model (followed by sampling and addition of i.i.d. Gaussian noise $\mathcal{N}(0,0.01)$). These are all essentially chaotic in nature (the added noise, in fact makes it stochastic). The goal is to multiplex all the four signals from neuron $Z$ to neuron $E$. In a similar fashion, the remaining $146$ neural signals $\{ s_i(n) \}_{i=5}^{150}$ are also obtained from the same model with the same settings of $r$ and $I$ but with randomly chosen initial conditions. As explained previously, the rest of the columns of $A$ (9850) are modeled as i.i.d Gaussian $\mathcal{N}(0,1)$ entries. Thus, we now have our ensemble of $10^4$ neural signals and neural noise which makes up Equation~(\ref{eqYequalAX}). For particular chosen values of the weights $ \{ w_i\}_{i=1}^{10^4}$, we plot $y(n)$ in Fig.~\ref{fig:ysignal} along with neural interference and noise.
%
%
%
%
%
%
%
\par Under normal circumstances, Equation~(\ref{eqYequalAX}) is unsolvable, unless $M \geq N$. This would imply that the neural signals need to be observed for a longer duration (thereby making $M$ at least $=10^4$). However, practical constraints such as immediate response to stimulus demand solving Equation~(\ref{eqYequalAX}) with shorter duration signals, which leads to the under-determined case ($M < N$). Surprisingly, this admits a unique solution which could be accurately determined via a recently developed signal processing paradigm known as compressed sensing (to be discussed in the next section), provided the $w_i$s to be recovered form a sparse-set (or a compressible set), and certain conditions  are satisfied by the matrix composed of the noisy neural signals.
\subsection{Compressed Sensing}
\label{subsec:cs}
\par Compressed sensing (CS) is a recent signal detection framework which has received considerable attention in the signal processing community~\cite{candes, eldar}. The signal measurement process is linear as in, the observed vector $y = Ax$. If the number of non zero entries of $x$ is less than or equal to $k$, we say $x$ is \emph{$k$-sparse}. The idea behind CS is that, if the received signal is sparse (in an appropriate domain), most of what is received is thrown away during compression or processing~\cite{introCS}. Thus, it is possible to reduce the number of measurements (as a function of the sparsity of $x$) and using random projections it is possible to recover the entire (sparse) signal exactly~\cite{meas}. Further, it has been shown that perfect recovery is possible in the presence of noise as well~\cite{candes}.

\par Let $x \in \mathbb{R}^N$ be a $k$-sparse signal. Let $A \in \mathbb{R}^{M \times N}$ denote the measurement matrix ($N=10^4$). The columns of the matrix correspond to the signals (vector-valued) generated by each of the $10^4$ neurons. The received vector $y \in \mathbb{R}^M$ is a linear combination of columns of $A$ where the coefficients form the vector $x$ i.e.,$y=Ax$.

\par In our problem of interest, this means that the signals from different neurons (columns of $A$) are multiplexed together during transmission to the nearby neuron. The number of signals multiplexed is equal to the sparsity of $x$.

\par Decoding of $x$ given $y$ can be carried out through a variety of techniques like greedy approach (algorithms like Orthogonal Matching Pursuit~\cite{tropp}) or optimization approach ($l_1$ minimization~\cite{l1min}), provided the measurement matrix $A$ satisfies a property commonly referred to as Restricted Isometry Property (RIP)~\cite{rip}. For example, when the entries are all drawn from a Gaussian distribution (say $\mathcal{N}(0,1)$) appropriately normalized, the matrix satisfies RIP. In these cases, it is possible to exactly recover $x$. However, in our problem, we are only interested in identifying the columns that contribute to the signal and not the exact amplitudes. In the CS parlance, this is called support recovery~\cite{support}.
\subsection{Neural Signal Demultiplexing: A Compressed Sensing Model}
\label{subsec:demux}
\par
\begin{figure}[!h]
\begin{center}
\centering
\resizebox{1.0\columnwidth}{!}{
\includegraphics{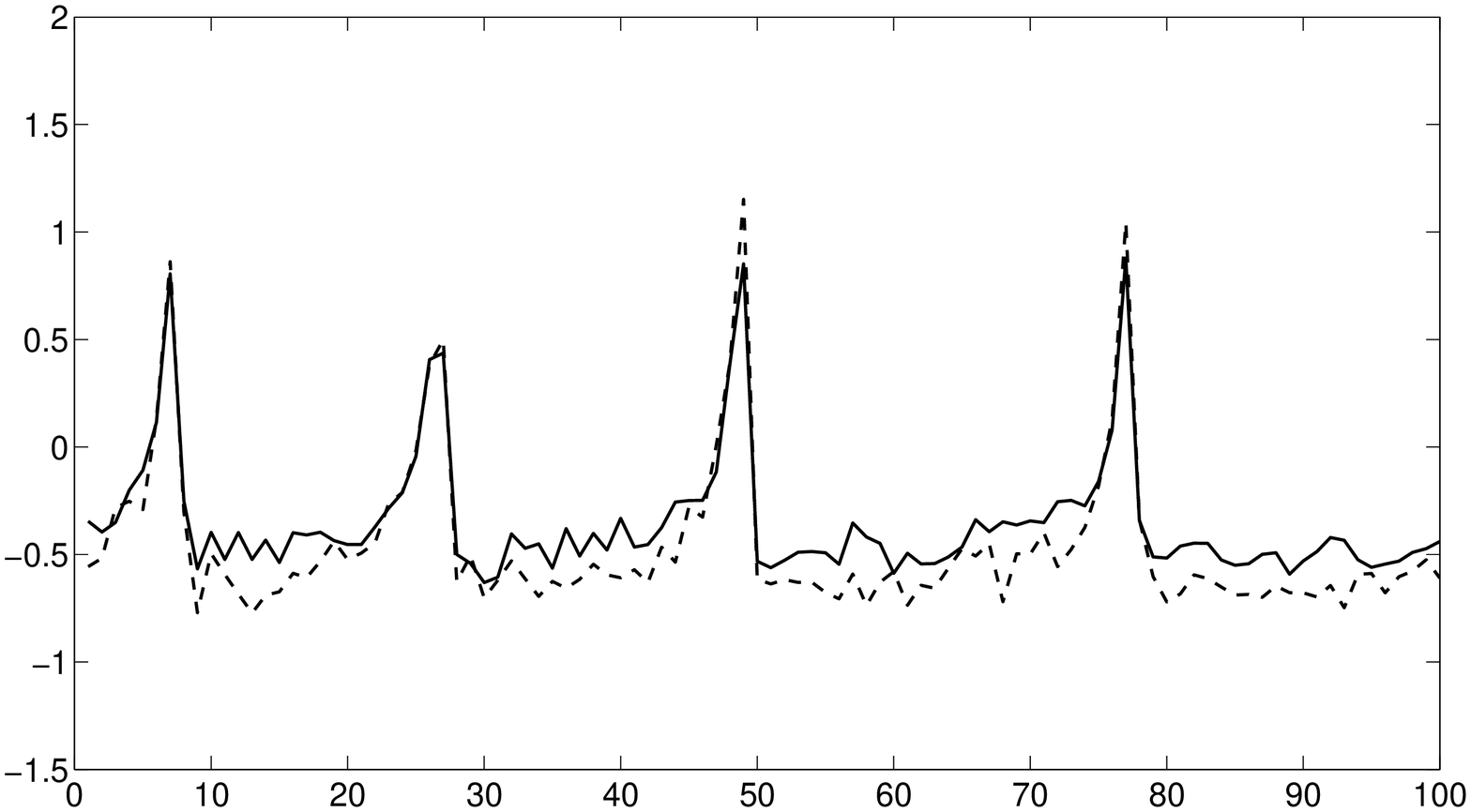}
}
\caption{Demultiplexing: Original signal $w_3 \cdot s_3$ (dotted) at neuron $Z$ and reconstructed signal $\hat{w_3} \cdot \hat{s_3}$ (bold) at neuron $E$ obtained by means of $l_1$ minimization. Similar recovery was obtained for $\hat{w_1} \cdot \hat{s_1}, \hat{w_2} \cdot \hat{s_2}$ and $\hat{w_4} \cdot \hat{s_4}$, but not shown here due to space constraints.} \label{fig:recon}
\end{center}
\end{figure}

 Consider the neuron $E$ which has received the signal $y=Ax$ from its neighbour $Z$. At $E$, it is only reasonable to assume that the signal set $A$ is not perfectly available. We consider the case where the neuron $E$ has to decode using a matrix $\hat{A}$ (instead of $A$) where noisy versions $1.5 \%$ of the columns of $A$ are present in $\hat{A}$ (these correspond to signals from nearby neurons, the signal-dominant set) and the rest $98.5 \%$ of the columns are composed of independently sampled Gaussian noise (these correspond to signals from far away neurons, the noise-dominant set). This could be mathematically represented as follows: Let $a_i, \hat{a}_i$ denote the $i^{th}$ columns of $A$ and $\hat{A}$ respectively. Consider a set of indices, corresponding to the signal-dominant columns of $A$ and let them be labelled $1,2,\ldots,150$. Then,
\begin{equation*}
\hat{a}_i = a_i + \epsilon_i,
\end{equation*}
where $\epsilon_i$s are independent Gaussian noise vectors.
Now, the problem is that of recovering $x$ given $\hat{A}$ and $y$ instead of $A$ and $y$. However, this is equivalent to decoding $x$ from $y=Ax+\epsilon$ where $\epsilon = \sum_{i=1}^{150} x_i\epsilon_i$ which has variance, say $\eta$. This can be achieved by solving the convex program
\begin{align*}
& \min ||x||_{l_1} &\text{ subject to }||Ax-y||_{l_2} \le \eta,
\end{align*}
which gives the sparse solution $x^*$. In fact, it is shown in~\cite{candes} that, under this setting,
\begin{equation*}
||x-x^*||_{l_2} \le C_1 \epsilon + C_2.\frac{||x-x_K||_{l_1}}{\sqrt{k}},
\end{equation*}
where $C_1$, $C_2$ are constants and $x_K$ is the same as $x$ at $k$ leading entries and the rest set to zero. Surprisingly, even in the absence of $98.5\%$ of columns (and the remaining columns corrupt), $l_1$ minimization is seen to perform well. This is evident from our simulation result depicted in Figure~\ref{fig:recon}. It shows that signals multiplexed and transported from $Z$ are successfully demultiplexed at $E$ despite having only a noisy, partial version of the mixing matrix $A$.
%
%
\subsection{Simulation settings}
\par We have used the following settings for our simulations. $N=10^4$, $M=100$, 150 signal-dominant neural signals from Hindmarsh-Rose model ($I = 3.28$, $r = 0.0021$ and randomly chosen initial conditions) with additive Gaussian noise $\mathcal{N}(0,0.01)$, 9850 noise-dominant neural signals simulated as additive Gaussian noise $\mathcal{N}(0,1)$, $k=4$ signals to be multiplexed with weights $w_1 = 1.0$, $w_2 = 0.9$, $w_3=0.8$ and $w_4=0.7$. The  weights for the signal-dominant columns $\{ w_i\}_{i=5}^{i=150}$ are drawn from uniform random distribution in $(0,0.02)$, weights for the noise-dominant columns $\{ w_i\}_{i=151}^{i=10^4}$ are drawn from uniform random distribution in $(0,0.001)$. CVX, a package for specifying and solving convex programs~\cite{cvx} was used for $l_1$ minimization. Threshold for choosing the $k$ significant weights (in $x^*$) was set at $T=0.4$.
\subsection{Discussion}
\label{subsec:disc}
\par It seems imperative to study the neural signal detection problem using the compressed sensing framework because it is one approach in which the number of measurements (size of the signal vector) is greatly reduced depending on how many signals are multiplexed. In the context of neural signal transmission, this means that the time to respond to a stimulus is reduced when fewer signals are multiplexed.

\par Usual decoding algorithms in CS aim at exact recovery of the vector $x$. However, in our problem, there are two aspects which makes decoding easier. Firstly, the neuron does not need the entire $x$, but only a few leading coefficients, which correspond to the signals of interest ($s_1(n)$, $s_2(n)$, $s_3(n)$ and $s_4(n)$). Secondly, the neuron does not need to extract the exact amplitude of these coefficients either, since it only needs to identify the ``active" columns of $A$ and no information is conveyed through spike amplitudes whatsoever. This corresponds to a support recovery problem and for certain classes of noise models ($l_\infty$-bounded, Gaussian), when the minimum amplitude is greater than $\sqrt{k}$ times the noise level where $k$ is the sparsity of $x$, exact recovery is possible~\cite{support}.

\par Besides, since CS delivers perfect recovery under noisy conditions as well~\cite{candes}, the recovery is robust even when the exact signal is not known at the receiver neuron ($E$ in our example) but only a noisy version like $\hat{A}$ in our example.

\par Another point of biological relevance is the connection between neural complexity ($N$), number of neural signals multiplexed ($k$) and the time to respond ($M$). Organisms with lower neural complexity, although have good response time, are likely to multiplex fewer signals. On the other hand, organisms with higher neural complexity (large $N$) can multiplex more signals and also have higher throughput of information transfer between adjacent neurons, albeit at the cost of longer response time. This relationship is very much in agreement with our proposed model since compressed sensing theory demands that the number of measurements $M$ obeys $M \approx C \cdot k \log \frac{N}{k}$~\cite{soda} for good recovery of $x$.
%
%
\section{Conclusions and Future Research Directions}
\label{conclusion}
\par The transport model between adjacent neurons in the brain that we have proposed has some unique features. The mixing matrix $A$ at pre-synaptic neuron $Z$ and the corresponding matrix $\hat{A}$ at the post-synaptic neuron $E$ need not have the same number of neural signals in the same order (number and position of columns can be different, and is most likely the case in the brain).  As a next step, it will be interesting to study the theoretical properties of the matrix $A$ (and $\hat{A}$) and the limits of such multiplexing. Experiments with real neural signals in the brain, to test the proposed model are necessary.
%
%


\begin{thebibliography}{1}

\bibitem{Azevedo2009}
F.~A.~Azevedo, L.~R.~Carvalho, L.~T.~Grinberg, J.~M.~Farfel, R.~E.~Ferretti, R.~E.~Leite, W.~Jacob Filho, R.~Lent, S.~Herculano-Houzel, ``Equal numbers of neuronal and nonneuronal cells make the human brain an isometrically scaled-up primate brain,'' \emph{J Comp Neurol.}, vol. 513, no. 5, pp. 532-541, Apr. 2009.


\bibitem{Meister2013}
M.~L.~R.~Meister, J.~A.~Hennig, A.~C.~Huk, ``Signal Multiplexing and Single-Neuron Computations in Lateral Intraparietal Area During Decision-Making,'' \emph{The J. of Neurosci.}, vol. 33, no. 6, pp. 2254 –2267, Feb. 2013.

\bibitem{Friedrich2004}
R.~W.~Friedrich, C.~J.~Habermann, G.~Laurent, ``Multiplexing using synchrony in the zebrafish olfactory bulb.,'' \emph{Nat. Neurosci.}, vol. 7, no. 8, pp. 862-871. Aug. 2004.

\bibitem{Panzeri2010}
S.~Panzeri, N.~Brunel, N.~K.~Logothetis, C.~Kayser, ``Sensory neural codes using multiplexed temporal scales,'' \emph{Trends Neurosci.}, vol. 33, no. 3, pp. 111-120, Mar. 2010.

\bibitem{Korn2003}
H.~Korn, P.~Faure, ``Is there chaos in the brain? II. Experimental evidence and related models,'' \emph{C. R. Biologies}, vol. 326, pp. 787–840, Sep. 2003.

\bibitem{Multiplex1}
L.~S. Tsimring, M.~M. Sushchik, ``Multiplexing chaotic signals using synchronization,'' \emph{Phys. Lett. A}, vol. 213, pp. 155--166, 1996.


\bibitem{Multiplex2}
Y.~Liu, P.~Davis, ``Dual synchronization of chaos,'' \emph{Phys. Rev. E}, vol.~61, pp. R2176--R2179, Mar. 2000.

\bibitem{Multiplex3}
S.~Sano, A.~Uchida, S.~Yoshimori, R.~Roy, ``Dual synchronization of chaos in Mackey-Glass electronic circuits with time-delayed
feedback,'' \emph{Phys. Rev. E}, vol.~75, no.~1,pp.~016207-1--016207-6, 2007.

\bibitem{Multiplex4}
J.~N.~Blakely, N.~J.~Corron, ``Multiplexing symbolic dynamics-based chaos communications using synchronization,'' in
\emph{J. Phys.: Conf. Ser.}, vol.~23, pp. 259--266, 2005.

\bibitem{Multiplex5}
N.~Nagaraj, P.~G.~Vaidya, ``Multiplexing of discrete chaotic signals in presence of noise,'' \emph{Chaos}, vol. 19, pp.~033102-1 - 033102-9, Jul. 2009.

\bibitem{Gupta2005}
P.~Gupta, D.~F.~Albeanu, U.~S.~Bhalla, ``Olfactory bulb coding of odors, mixtures and sniffs is a linear sum of odor time profiles,'' \emph{Nat. Neurosci.}, vol.~18, no.~2, pp.~272-281, Feb. 2015.

\bibitem{Hindmarsh1984}
J.~L.~Hindmarsh, R.~M.~Rose, ``A model of neuronal bursting using three coupled first order differential equations,'' \emph{Proc. R. Soc. London, Ser. B, Bio. Sci.}, vol.~221, no.~1222, pp.87-102, Mar. 1984.

\bibitem{Rabinovich1997}
M.~I.~Rabinovich, H.~D.~I.~Abarbanel, R.~Huerta, R.~Elson, A.~I.~Selverston, ``Self-regularization of chaos in neural systems: Experimental and theoretical results,'' \emph{IEEE Trans. Cir. Sys. I,}  vol.~44, no.~N10, pp.~997-1005, Oct. 1997.

\bibitem{Gonzalez2003}
J.~M.~Gonz\'{a}lez-Miranda, ``Observation of a continuous interior crisis in the Hindmarsh–Rose neuron model,'' \emph{Chaos}, vol.~13, no.~3, pp.~845-852, Aug. 2003.

\bibitem{candes}
E.~J.~Candes, J.~Romberg, T.~Tao, ``Stable signal recovery from incomplete and inaccurate measurements", \emph{Comm. Pure Appl. Math.}, 59: 1207–1223, 2006.

\bibitem{eldar}
M.~F.~Duarte, Y.~C.~Eldar,``Structured Compressed Sensing: From Theory to Applications", \emph{IEEE Transactions on Signal Processing}, vol.59, no.9, pp.4053-4085, Sept. 2011.

\bibitem{introCS}
E.~J.~Candes, M.~B.~Wakin, "An Introduction To Compressive Sampling," \emph{IEEE Signal Processing Magazine}, vol.25, no.2, pp.21-30, March 2008.

\bibitem{meas}
R.~Baraniuk, M.~Davenport, R.~DeVore, M.~B.~Wakin, ``A Simple Proof of the Restricted Isometry Property for Random Matrices", \emph{Constructive Approximation}, vol 28, pp. 253-263, 2008.

\bibitem{tropp}
J.~A.~Tropp, A.~C.~Gilbert, ``Signal Recovery From Random Measurements Via Orthogonal Matching Pursuit", \emph{IEEE Transactions on Information Theory},  vol.53, no.12, pp.4655-4666, Dec. 2007.

\bibitem{l1min}
E.~J.~Candes, J.~Romberg, T.~Tao, ``Robust uncertainty principles: exact signal reconstruction from highly incomplete frequency information," \emph{IEEE Transactions on Information Theory}, vol.52, no.2, pp.489-509, Feb. 2006.

\bibitem{rip}
E.~J.~Candes, ``The restricted isometry property and its implications for compressed sensing", \emph{Comptes Rendus Mathematique}, Volume 346, Issues 9–10, pp 589-592, May 2008.

\bibitem{support}
R.~Wu, W.~Huang, D.~-R.~Chen, ``The Exact Support Recovery of Sparse Signals With Noise via Orthogonal Matching Pursuit", \emph{IEEE Signal Processing Letters}, vol.20, no.4, pp.403-406, April 2013.

\bibitem{cvx}
M.~Grant, S.~Boyd, ``CVX: Matlab software for disciplined convex programming", version 2.0 beta. http://cvxr.com/cvx, September 2013.

\bibitem{soda}
K.~D.~Ba, P.~Indyk, E.~Price, D.~P.~Woodruff, ``Lower bounds for sparse recovery", \emph{Proceedings of the twenty-first annual ACM-SIAM symposium on Discrete Algorithms (SODA '10)}, pp. 1190-1197, 2010.


%
%
%
%
\end{thebibliography}
\end{document}